\documentclass[]{jfm}

\usepackage{graphicx}
\usepackage{newtxtext}
\usepackage{newtxmath}
\usepackage{natbib}
\usepackage{hyperref}
\hypersetup{
    colorlinks = true,
    urlcolor   = blue,
    citecolor  = black,
}

\newcommand{\RomanNumeralCaps}[1]
\title{Thermodynamic growth of sea ice: assessing the role of salinity using a quasi-static modelling framework}

\author{David W. Rees Jones\aff{1}
  \corresp{\email{david.reesjones@st-andrews.ac.uk}}
  }

\affiliation{\aff{1} School of Mathematics and Statistics, University of St Andrews, North Haugh, St Andrews, KY16 9SS}

\begin{document}
\maketitle

\begin{abstract}
Sea ice is a mushy layer, a porous material whose properties depend on the relative proportions of solid and liquid. The growth of sea ice is governed by heat transfer through the ice together with appropriate boundary conditions at the interfaces with the atmosphere and ocean. The salinity of sea ice has a large effect on its thermal properties so might naively be expected to have a large effect on its growth rate. However, previous studies observed a low sensitivity throughout the winter growth season. The goal of this study is to identify the controlling physical mechanisms that explain this observation. We develop a simplified quasi-static framework by applying a similarity transformation to the underlying heat equation and neglecting the explicit time dependence. We find three key processes controlling the sensitivity of growth rate to salinity. First, the trade-off between thermal conductivity and (latent) heat capacity leads to low sensitivity to salinity even at moderately high salinity and brine volume fraction. Second, the feedback on the temperature profile reduces the sensitivity relative to models that assume a linear profile, such as zero-layer Semtner models. Third, thicker ice has the opposite sensitivity of growth rate to salinity compared to thinner ice, sensitivities that counteract each other as the ice grows. Beyond its use in diagnosing these sensitivities, we show that the quasi-static approach offers a valuable sea-ice model of intermediate complexity between zero-layer Semtner models and full partial-differential-equation-based models such as Maykut--Untersteiner/Bitz--Lipscomb and mushy-layer models.  
\end{abstract}


\section{Introduction}
\label{sec:intro}
Sea ice is a fundamental part of the climate of the polar regions. Sea ice evolves as it is transported by wind and ocean currents, due to various mechanical processes such as ridging, and due to thermodynamic processes of growth and melting \citep{Golden2020}. This study focuses on the thermodynamic aspects of the evolution. 

The first goal of this study is to understand the effect of salinity on sea-ice growth through the winter growth season and to compare different representations of this process in active use in the sea-ice modelling community (these include zero-layer Semtner, Maykut--Untersteiner/Bitz--Lipscomb and dynamic-salinity mushy-layer models). The second goal is to propose a new type of quasi-static approximate model of an intermediate complexity between a zero-layer model (based on ordinary differential equations) and the other types of models which are all based on partial differential equations. 

The classic early work on sea-ice growth was developed by Stefan in the late nineteenth century \citep{Stefan1891}. Historical reviews of Stefan's contribution are given by \citet{Vuik1993} and \citet{Sarler1995}. Stefan investigated solidification problems with a free boundary between the solid and liquid phases, a class of problems subsequently called \textit{Stefan problems}. Stefan introduced several approaches that will guide this study. The growth of ice is governed by a heat equation. Stefan applied a similarity-solution method to solve this equation analytically and also obtained approximate solutions under the assumption of a linear temperature profile. In either case, the ice thickness increases with the square root of time. This theoretical prediction was consistent with measurements of sea-ice thickness from polar expeditions. 

Stefan treated sea ice as a pure material with constant thermal properties. However, sea ice forms from seawater, an alloy of water and salt (of various types). The salt retained in sea ice allows pockets of saltwater (brine) to remain as liquid inclusions within a matrix of solid ice even when the composite is cooled beneath the freezing point of seawater. This part-solid and part-liquid composite is called a mushy layer. \citet{Huppert1985} and \citet{Worster1986} 
extended the similarity solution for a pure material to the solidification of  a two-component (binary) alloy. A two-component alloy is a simple model of saltwater (amongst many other chemical systems). As for the pure system, the solidification front advances with the square root of time. In this study, we develop a new quasi-static approximation. This approximation reduces to the similarity-solution method under certain conditions and so represents an extension of this body of earlier work as well as Stefan's original application of the idea. 

\citet{Kerr1989,Kerr1990} extended these earlier models to account for the variation of the thermal properties within the mushy layer and for variation in the bulk composition within the mushy layer. These papers parameterized heat transfer through the liquid as a turbulent heat flux rather than assuming that all the heat transfer was conductive. These extensions are important for modelling sea-ice growth because the thermal properties vary considerably and also because the heat transfer from the ocean is dominantly through turbulent convection rather than conduction. The resulting equations were solved numerically. However, the treatment was still limited because it was not known how to calculate the evolution of the bulk composition (salinity) within the mush. In practice, sea ice is observed to desalinate rapidly through a process called convective desalination driven by a compositional buoyancy gradient within the ice \citep[see reviews by][]{Worster1997,Worster2000,Anderson2020}. 

Within the sea-ice literature, \citet{Maykut1969,Maykut1971} developed a significant new type of model (see references therein for some antecedent developments). We will refer to this model by the acronym MU. There are some similarities with the developments in mushy-layer theory outlined above (which all came later), particularly in terms of the role of salinity in the thermal properties of the ice. The MU model has the same limitation as the mushy-layer models mentioned so far in terms of having to prescribe the salinity. There are also differences. In particular, the ice is topped by a layer of snow, radiative heating is included and a realistic energy balance with the atmosphere is used. The MU model is of particular importance because it is still widely used in large-scale climate simulations, particularly in the version developed further by \citet{Bitz1999}. \citet{Bitz1999} updated the MU model by accounting for the latent heat of melting the upper surface of the ice, which is important for modelling the summer melt season. However, for the winter growth season, this difference is not important since there is limited melting at that interface in winter. 

The thermodynamics within the sea ice itself in the MU model is structurally entirely consistent with that used in mushy-layer theory \citep{Feltham2006}. Indeed, by a suitable choice of the parameter values, they can be made identical. However, the parameters used by default in MU models have a stronger sensitivity to salinity than would be suggested by mushy-layer theory (in which the thermal conductivity is a weighted average of the solid and liquid conductivity). We will return to this issue later when considering suitable parameter values.

Another widely used class of sea-ice models was developed by \citet{Semtner1976}. 
These models use the same thermodynamic framework as the MU model, but rather than solving the full heat equation, the equation is vertically discretized into  three layers (one snow layer and two sea-ice layers), with the temperature field assumed linear between the midpoints of each layer. In an appendix,  \citet{Semtner1976} proposed an even simpler `zero-layer' model, in which the heat equation is not solved at all and the temperature field varies linearly across the ice. As mentioned above, this type of linear approximation was first used by \citet{Stefan1891}. We will discuss why such approximations continue to work well even when the thermal properties vary strongly across the ice.

Recently, a new generation of dynamic-salinity mushy-layer models has been developed \citep{Turner2013,Griewank2013,ReesJones2014,Griewank2015}. These applied theoretical developments within mushy-layer theory to calculate the evolution of bulk salinity either using a semi-analytical theory \citep{ReesJones2013a,ReesJones2013b} or two-dimensional numerical simulations \citep{Wells2011,Wells2013}. The particular implementation of \citet{Turner2013} has been implemented in large-scale models as discussed below. However, the three dynamic salinity models are all rather similar from a theoretical point of view \citep{Worster2015}. Laboratory experiments of the very early stages of ice growth by \citet{Thomas2020} suggest that dynamic-salinity mushy-layer models can successfully capture the initial desalination of sea ice and perform better than other types of parameterization. 

Indeed, collectively the zero-layer, MU and dynamic-salinity mushy-layer models represent the three options available to users of the CICE/Icepack software packages \citep{CICE2024,icepack2024}. These packages are widely used sea-ice models for large-scale simulations. A recent inter-comparison of CMIP6 models shows that these three models of sea-ice thermodynamics remain in use, with the Bitz--Lipscomb version of the MU model being the most common \citep{Keen2021}. One goal of our study is to understand better the differences between such models. This is motivated in part by numerical calculations in \citet{Griewank2013} and \citet{ReesJones2014} that suggested introducing a dynamic salinity field had only a small effect on the ice growth. Similar results had previously been obtained with different prescribed salinity profiles by \citet{Vancoppenolle2005}. These observations are somewhat curious given the large variation in thermal properties through sea ice and how strongly these depend on salinity. In this study, we will provide detailed physical reasons for this observation and show that it continues to hold across a wide range of sea-ice salinities (from ice that is almost as salty as seawater to ice that is completely desalinated). 

Our study also contributes a new quasi-static approximate model of sea ice which has an intermediate complexity between zero-layer models and partial-differential-equation-based models (such as MU-type and mushy-layer models). Under constant boundary conditions, this model reduces to the similarity-solution method mentioned previously and gives an exact solution to MU-type models. Under variable boundary conditions, the solution is only approximate and so we investigate the validity of the approximation on theoretical grounds and through numerical calculation. While our study focuses on terrestrial sea ice, this type of intermediate complexity approach may be attractive for emerging research into the postulated mushy layers on icy moons \citep[e.g.,][]{Buffo2020,Buffo2021,Vance2021}. 

The structure of the paper is as follows. 
In section~\ref{sec:method}, we develop governing equations, non-dimensionalize them, discuss appropriate parameter values, and develop the quasi-static approximation. 
In section~\ref{sec:initial}, we analyze the initial growth rate of sea ice using theoretical (asymptotic) analysis and numerical calculation. We explain why salinity has a very weak effect on the growth rate. We compare our calculations to those based on zero-layer and MU-type models. 
In section~\ref{sec:later-stage}, we consider the subsequent growth rate, which diminishes as the ice grows and the ocean heat flux becomes more important. We calculate the evolution of ice thickness and develop an approximate analytical solution. We return to the question of the role of salinity on ice growth. 
In section~\ref{sec:time-dependent}, we consider various kinds of variable conditions, particularly time-dependent atmospheric temperature and time-dependent sea-ice salinity. Then we test the validity of the quasi-static approximation by comparing it against numerical solutions of a full partial-differential-equation-based model. 
Finally, in section~\ref{sec:implications}, we summarize our findings from the viewpoint of the implications for large-scale sea-ice modelling.  

\section{Model of sea-ice growth and the quasi-static approximation}\label{sec:method}
Sea ice is formed from seawater and consists of solid ice (the \textit{solid phase}) and liquid saltwater/brine (the \textit{liquid phase}). The composite will be referred to as \textit{sea ice}, or \textit{ice} for brevity. 

The thermodynamic growth of ice is a predominantly one-dimensional process in which ice grows in the vertical $z$ direction, where $z$ increases downwards. The top of the ice (at the boundary with the atmosphere) lies at $z=0$ and the bottom (at the boundary with the ocean) lies at $z=h$, where $h$ is the ice thickness. The thickness evolves in time $t$ so $h=h(t)$. The primary model output is the growth rate $\dot{h}(t)$ and hence, by integrating in time, the thickness $h(t)$.

A key simplifying assumption in our model is that we treat the bulk salinity $S$ as constant. Correspondingly, we neglect the effect of brine advection through the pores, which is the primary mechanism behind salinity evolution. As discussed in section~\ref{sec:intro}, salinity evolution is an important process. However, we wish to understand why the salinity only has a small effect on ice growth, so it is simpler to have a single constant characterizing the salinity.  
In appendix~\ref{app:heat}, we show that both the direct advective transport of heat by brine advection and the latent heat changes associated with changes in bulk salinity can be self-consistently neglected. 
Moreover, taking the bulk salinity as constant satisfies local salt conservation in the absence of advection and diffusion [equation (4) in \citet{ReesJones2014}]. 
It also satisfies global salt conservation: there is a salt flux proportional to the ice growth rate and the difference in bulk salinity across the ice--ocean interface [equation (18) in \citet{ReesJones2014}].
While the more complete descriptions of salinity evolution described in section~\ref{sec:intro} are more realistic; nevertheless, assuming constant salinity does satisfy local and global salt conservation.
Thus we can safely analyze the temperature evolution while holding the salinity constant. 

\subsection{Temperature equation within the ice}
The growth rate of ice depends on thermal transfer through the ice governed by conservation of heat within the ice,
\begin{equation} \label{eq:temp_0}
\overline{c} \frac{\partial T}{\partial t} =   \frac{\partial}{\partial z} \left(\overline{k} \frac{\partial T}{\partial z} \right)- L \frac{\partial X}{\partial t},
\end{equation}
where $T(z,t)$ is the temperature and $X(z,t)$ is the liquid (brine) fraction. The remaining quantities are the thermal properties: $L$ is the volumetric latent heat, $\overline{c}$ is the volumetric heat capacity and $\overline{k}$ is the thermal conductivity. The properties $\overline{c}$ and $\overline{k}$ depend on the relative fraction of the ice that is occupied by solid and liquid. Weighting these properties volumetrically,
  \begin{subequations} \begin{align} \label{eq:kc-def}
 \overline{k} & = 
     k_lX + k_s(1-X) \equiv k_s\left[1-X \Delta k  \right],\\[3pt]
   \overline{c} & = 
     c_lX + c_s(1-X) \equiv c_s\left[1-X \Delta c  \right],
\end{align} \end{subequations}  
where a subscript $s$ denotes a property of the solid phase and $l$ denotes a property of the liquid phase. For the final equivalences, we define
  \begin{subequations} \begin{align} \label{eq:Delta-kc-def}
 \Delta k &= (k_s-k_l)/k_s,\\[3pt]
   \Delta c &= (c_s-c_l)/c_s.
\end{align} \end{subequations}  
These are dimensionless measures of the differences between the thermal properties of solid and liquid phases.

The liquid fraction depends on the temperature of the ice as well as its bulk salinity $S$, i.e., the weighted average of the salinity of the solid phase (which is assumed to be zero) and that of the liquid phase $C$. These are related by 
\begin{equation} \label{eq:bulk-salinity-def}
S=CX.
\end{equation}
We assume that phase change occurs until the system reaches a local thermodynamic equilibrium in which the liquid phase lies at the temperature-dependent freezing point. Thus 
\begin{equation} \label{eq:C(T)}
C=-T/m,
\end{equation}
where $m$ is the slope of the freezing point (assumed constant). Hence, by combining with equation~\eqref{eq:bulk-salinity-def}, we determine the liquid fraction
\begin{equation} \label{eq:liq-fraction}
X = \frac{mS}{-T},
\end{equation}
and by the chain rule
\begin{equation} \label{eq:temp_X}
\frac{\partial X}{\partial t} = X\frac{1}{-T}\frac{\partial T}{\partial t} .
\end{equation}

Equation~\eqref{eq:temp_X} allows the latent heat term on the RHS of equation~\eqref{eq:temp_0} to be rewritten as an enhanced heat capacity. In particular, we write
\begin{equation} \label{eq:temp_1}
c \frac{\partial T}{\partial t} =  \kappa_s \frac{\partial}{\partial z} \left(k \frac{\partial T}{\partial z} \right),
\end{equation}
where $\kappa_s=k_s/c_s$ is the thermal diffusivity of the solid phase and
  \begin{subequations} \begin{align}
k& = 
     1-X \Delta k  ,\\[3pt]
   c & = 
    1-X \Delta c +\frac{L}{c_s(-T)} X ,    
\end{align}  \end{subequations}  
are the dimensionless thermal conductivity and heat capacity respectively. These quantities depend on $T$, so equation~\eqref{eq:temp_1} is a nonlinear diffusion equation. 

\subsection{Boundary conditions}
In general, suitable boundary conditions for sea-ice evolution require the conservation of heat and salt across the interfaces. Here, we take a simplified approach. 

We assume that the top of the ice is held at some temperature $T_B$, the atmospheric temperature. In principle, this temperature could vary in time, in which case $T_B$ would denote the mean (time-averaged) or initial atmospheric temperature. 
In mushy-layer models, such a fixed temperature boundary condition is commonly used and corresponds physically to a perfectly conducting boundary. In sea-ice models, an energy flux balance is usually used. \citet{Hitchen2016} showed that such flux balance models can be linearized and expressed as a Robin-type boundary condition involving both $T$ and $\partial T/\partial z$; such boundary conditions could be handled within our modelling framework. 

We assume that the bottom of the ice is at the freezing point of seawater $T_0= -m S_0$, where $S_0$ is the salinity of seawater. This introduces a natural temperature scale to the problem 
\begin{equation}
\Delta T \equiv T_0-T_B,
\end{equation}
which will be positive for ice growth. 

Although equation~\eqref{eq:temp_1} is a second-order equation and we already have two boundary conditions, we need an extra boundary condition to determine the evolution of the free boundary $h(t)$. Its evolution is determined by a generalized Stefan condition \citep{Kerr1989,Kerr1990}. This condition represents the balance of heat fluxes across a narrow boundary layer below the ice--ocean interface
\begin{equation} \label{eq:bc_1}
\dot{h}\left[c_l(T_l-T_0)\ + L (1-X)_{z=h^-}\right] +  F_T = k_s \left(k\frac{\partial T}{\partial z}\right)_{z=h^-},
\end{equation}
where 
$T_l>T_0$ is the temperature of the upper ocean (assumed constant) and $F_T$ is the turbulent heat flux supplied by the ocean. In general, the LHS of equation~\eqref{eq:bc_1} is dominated by the latent heat term, but we retain the sensible heat term $c_l(T_l-T_0)$ as it regularises the boundary condition when the liquid fraction at the interface $X_{z=h^-} =1$. 

We introduce a scale for the bulk salinity of sea ice relative to the salinity of the ocean
\begin{equation} \label{eq:hatS_def}
\hat{S}=S/S_0,
\end{equation}
which will lie between 0 and 1 depending on how much the sea ice has desalinated (we discuss the range of $\hat{S}$ in section~\ref{sec:parameters}). Note that $X_{z=h^-} = \hat{S}$, so $\hat{S}$ can also be thought of as a scale for the liquid fraction.  

\subsection{Non-dimensionalization}
The growth of ice appears to have no natural length scale. Previous studies of growth in the laboratory have typically used the depth of the tank as a scale \citep[e.g.,][]{Kerr1990}. However, if we are interested in the long-term growth of ice, it makes more sense to use the following scale  
\begin{equation}
h_\infty = k_s \frac{\Delta T}{F_T},
\end{equation}
which is a scale estimate for the steady-state ice thickness based on estimating the steady-state balance in equation~\eqref{eq:bc_1}. It is important to note that this is a scale estimate, not the actual steady-state thickness. This scale is chosen because we want to investigate how ice growth depends on salinity, so it is convenient to non-dimensionalize with respect to a scale that does not depend on salinity. We report the actual steady-state thickness of the model in section~\ref{sec:equil}.

Thus we introduce a rescaled height ($\hat{h}$) as well as a rescaled time ($\tau$) and distance ($\zeta$) coordinate system:
\begin{equation}
\hat{h}\equiv \frac{h}{h_\infty}, \qquad \tau \equiv \frac{t \kappa_s}{h_\infty^2}, \qquad \zeta \equiv \frac{z}{h_\infty \hat{h}}. 
\end{equation}
The scaling for time is based on thermal diffusion across the steady-state ice thickness.
The scaled temperature of ice can be written 
\begin{equation}
\theta=\frac{T-T_B}{\Delta T},
\end{equation}
such that $\theta$ runs between 0 and 1 from the upper (atmospheric) to the lower (oceanic) interface.
We denote the scaled cold atmospheric temperature  
\begin{equation}
\theta_B=\frac{-T_B}{\Delta T},
\end{equation}
which is greater than 1 by the definition of $\Delta T$. 
The effective scaled temperature difference across the ice--ocean boundary layer is
\begin{equation}
\theta_e=\frac{c_l\left(T_l-T_0\right)}{L},
\end{equation}
where we non-dimensionalize with respect to latent heat because this ratio controls the relative importance of sensible to latent heat fluxes in equation~\eqref{eq:bc_1}.

The Stefan condition, equation~\eqref{eq:bc_1}, can be rescaled
\begin{equation} \label{eq:bc_nd1}
\hat{h} \frac{d\hat{h}}{d\tau} = \frac{1}{\hat{L}} q,
\end{equation}
where the scaled latent heat $\hat{L}=L/(c_s\Delta T)$ is often called the Stefan number, and  the growth rate factor $q$ is defined by 
\begin{equation} \label{eq:bc_nd1_qdef}
q = \frac{ \left.k\frac{\partial \theta}{\partial \zeta}\right|_{\zeta=1^-} -\hat{h}}{1-\hat{S}+\theta_e}.
\end{equation}
Thus,
\begin{equation} \label{eq:q_dh}
 \frac{d}{d\tau}\left(\frac{\hat{h}^2 \hat{L}}{2}\right)=q,
\end{equation}
so determining $q$ tells us how fast the square of ice thickness changes. 
We will often refer to $q$, which is the crucial output of our calculations, as the growth rate factor. 

Similarly, the temperature equation~\eqref{eq:temp_1} can be written in scaled form
\begin{equation} \label{eq:temp_nd1}
c \hat{h}^2 \frac{\partial \theta}{\partial \tau} - \frac{c \zeta q}{\hat{L}} \frac{\partial \theta}{\partial \zeta}  =   \frac{\partial}{\partial \zeta} \left(k \frac{\partial \theta}{\partial \zeta} \right),
\end{equation}
where scaled versions of the material properties are given by   \begin{subequations} \begin{align}
X& = 
     \hat{S}\frac{\theta_B-1}{\theta_B-\theta}  ,\\[3pt]
     k& = 
     1-X \Delta k  ,\\[3pt]
   c & = 
    1-X \Delta c +\frac{\hat{L}}{\theta_B-\theta} X.     \label{eq:c_ndim}
\end{align} \end{subequations}  
Although equation~\eqref{eq:temp_nd1} does not include any direct advective transport, the second term on the LHS is a pseudo-advection term proportional to $q$ associated with the changing domain.

\subsection{Parameter values and approximations} \label{sec:parameters}
Six dimensionless parameters characterize the problem. In this section, we give typical values or ranges for these parameters and discuss appropriate approximations. We also recast two parameters to better separate material properties (that are essentially fixed) from quantities that vary depending on environmental conditions. Throughout this section, we use material properties given in table 1 of \citet{ReesJones2014}; see references therein for further discussion.

First, the difference between solid and liquid conductivities $\Delta k =1-k_l/k_s\approx 0.76$. This means that the thermal conductivity of ice is about 4 times larger than that of water, a significant difference. 
However, various MU-type models use a formula for the thermal conductivity of the form
 
\begin{equation*}
\overline{k}=k_s + \beta S /T,
\end{equation*}
 
where $\beta$ is an empirical constant. By comparing this expression with equations~(\ref{eq:kc-def}\textit{a}) and \eqref{eq:liq-fraction}, we see that the MU-type models are equivalent provided $\Delta k =\beta /m k_s$. Particular choices of parameters differ slightly between publications, so, as an example, we consider the default parameter values given in the CICE/Icepack documentation \citep{CICE2024,icepack2024}. These list $\beta=0.13$~W/m/ppt, $m= 0.054$~deg/ppt and $k_s= 2.03$ W/m/deg, which combine to give $\Delta k \approx 1.2$. This is physically problematic in the framework of mushy-layer theory because $\Delta k \leq 1 $ by definition~(\ref{eq:Delta-kc-def}a). Indeed, the Icepack documentation casts doubt on the suitability of these parameter values based on experimental results \citep{icepack2024}. However, it appears not widely known that these default parameter values are inconsistent with mushy-layer theory. We comment more generally on this issue in the context of large-scale models in section~\ref{sec:implications-large-scale}. 

Second, the difference between solid and liquid heat capacities $\Delta c =1-c_l/c_s\approx -1.1$. Equivalently, the heat capacity of water is about double that of ice. This is a significant difference. However, by inspecting equation~(\ref{eq:c_ndim}\textit{c}), we see that the magnitude of $\Delta c$ should be compared with the role of latent heat since both appear multiplied by $X$. Indeed the ratio of the second and third terms is
\begin{equation} \label{eq:delta_c_ratio}
\frac{\Delta c}{{\hat{L}}/
({\theta_B-\theta})} \leq \frac{\Delta c \, \theta_B}{\hat{L}} =\frac{c_s-c_l}{L} (-T_B),
\end{equation}
where the inequality arises from the fact that $\theta>0$. Even for quite strong cooling, e.g., \mbox{$T_B=-20\, ^\circ$C}, the ratio in equation~\eqref{eq:delta_c_ratio} is $\lesssim 0.13$. 
For numerical calculations, we will retain the parameter $\Delta c$. However, it only has a small effect on results and we will neglect it when carrying out analysis to reduce the number of parameters. 

Third, the effective ice--ocean temperature difference $\theta_e=c_l\left(T_l-T_0\right)/L\ll 1$, because \mbox{$L/c_l \approx 77 \, ^\circ$C} and any temperature difference is typically less than a degree (and often much smaller). In some mushy-layer models of sea ice, this term still plays an important role, because it regularizes the Stefan condition in the case that the liquid fraction $X=1$ at the interface (equation~\ref{eq:bc_1}). However, in our model with a fixed sea-ice salinity, it is reasonable to neglect $\theta_e$ provided $\hat{S}$ is not very close to 1 (which ensures the liquid fraction is not 1 at the interface). Formally, we need $\theta_e \ll 1-\hat{S}$. For numerical calculations, we will use the representative value \mbox{$T_l-T_0\approx 0.017 \, ^\circ$C}, which gives \mbox{$\theta_e \approx 2\times 10^{-4}$}. For analytical calculations, we neglect this term as an excellent approximation. 

Fourth, the effective latent heat or Stefan number $\hat{L}=L/(c_s\Delta T) \gg 1$, because  \mbox{$L/c_s \approx 160\, ^\circ$C} which is much greater than a typical temperature difference across the ice. For example, if $\Delta T \approx 20\, ^\circ$C, then $\hat{L}\approx 8$. The definition of $\hat{L}$ is convenient for simplifying and analyzing the equations. However, one limitation of this formulation is that changing the cold atmospheric temperature $T_B$, which will vary depending on the environmental conditions, changes both $\hat{L}$ and $\theta_B$. Therefore, we introduce an alternative effective latent heat 
\begin{equation}
\hat{L}_0 = \frac{L}{c_s (-T_0)} \approx 83,
\end{equation}
which, to an excellent approximation, is a fixed material property, because the freezing point of seawater is roughly constant (assuming its salinity does not vary much).

Fifth, the scaled cold atmospheric temperature $\theta_B=-T_B/\Delta T>1$, which ensures that the atmospheric temperature is below the freezing point $T_0$. As $T_B$ approaches $T_0$, the temperature differences $\Delta T$ approaches zero, so $\theta_B$ can be arbitrarily large. So we introduce an alternative parameter
\begin{equation} \label{eq:theta_0}
\theta_0=\left(\theta_B-1\right)^{-1} \equiv \frac{\Delta T}{-T_0}, \quad \Leftrightarrow \quad \theta_B=1+\theta_0^{-1},
\end{equation}
where the equivalence expresses $\theta_0$ in terms of dimensional quantities. In a similar way to $\hat{L}_0$, it is convenient to scale against a fixed quantity. Thus $\theta_0$ varies between 0 (when there is no freezing) and about 9.4 (when \mbox{$T_B=-20\, ^\circ$C} and there is strong freezing). Unless otherwise stated, we take a default value $\theta_0\approx 9.4$. Then the Stefan number can be rewritten
\begin{equation} \label{eq:Lhat-L0}
\hat{L}=\hat{L}_0 \theta_0^{-1}.
\end{equation}
Thus, variation in $\theta_0$ should be interpreted as representing variation in the environmental conditions, which in turn controls variation in $\hat{L}$. Combining the default values for $\hat{L}_0$ and $\theta_0$ gives a default value for $\hat{L}\approx 8.3$.

Sixth, the sea-ice salinity $\hat{S}$ defined in equation~\eqref{eq:hatS_def} will lie between 0 and 1. In mushy-layer theory, the salinity is constant across the ice-ocean interface, so $\hat{S}=1$ at the interface. Equivalently, all the salt contained in seawater is initially incorporated into the sea ice as liquid brine inclusions \citep{Notz2008}. Some laboratory experiments suggest that there is a delay of several hours before desalination begins \citep{Wettlaufer1997JFM}. However, this is a relatively short time period and such a delay was not apparent in the field observations of \citet{Notz2008}. Within about 12 hours, sea ice appears to lose at least half the salt originally contained in seawater \citep{Notz2008,Thomas2020} so for all but the very initial stages of ice growth, it is reasonable to take $\hat{S}\lesssim 0.5$. For our results, we consider salinities between $\hat{S}=0$ (fresh ice) and $\hat{S}=0.8$ (very salty ice, in dimensional units about 28~ppt, which is much saltier than even very young ice is observed to be) to consider a very broad possible range.

\begin{figure}
  \centerline{\includegraphics[width=1.0\textwidth]{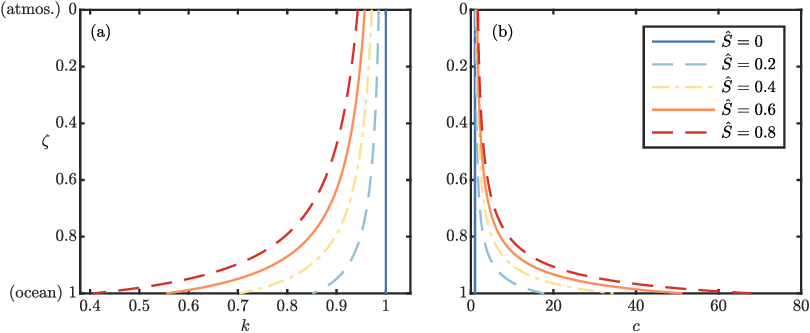}}
  \caption{Depth dependence of the thermal properties of sea ice from atmosphere ($\zeta=0$) to ocean ($\zeta=1$). (a) The thermal conductivity and (b) the heat capacity of sea ice vary considerably with salinity $\hat{S}$. The depth dependence was calculated by assuming that temperature varied linearly with depth ($\theta=\zeta$) in equation~\ref{eq:c_ndim}.}
\label{fig:summary}
\end{figure}

The values of $\Delta k$ and $\Delta c$ ensure that the effective conductivity and heat capacity of sea ice vary considerably in sea ice (figure~\ref{fig:summary}). Saltier ice is less thermally conductive because the liquid fraction is higher. However, saltier ice has a higher heat capacity. The fact that the heat capacity is much greater than 1 reflects the fact that the heat capacity (\ref{eq:c_ndim}\textit{c}) is dominated by latent heat release. 

\subsection{Dimensional parameter values}\label{sec:par-dim}
The solution of the dimensionless problem only involves the dimensionless parameters introduced previously. However, to convert back to dimensional form, we additionally need the dimensional steady-state thickness estimate $h_\infty$.

In addition to the material properties given in table 1 of \citet{ReesJones2014}, we need to specify a value of $F_T$, $T_0$ and $T_B$. We take \mbox{$F_T=0.0013$~J/s/cm$^2$} and \mbox{$T_0=-1.9\,^\circ$C}. This value of $F_T$ is smaller than (about half) that used in \citet{ReesJones2014} and was chosen to achieve a sensible equilibrium thickness, comparable to \citet{Maykut1971}. The value of $F_T$ used is within the range observed by \citet{Wettlaufer1991}. For $T_B$, we consider two possible values: 
\begin{enumerate}[i]
\item
For \mbox{$T_B=-20 \,^\circ$C}, we obtain \mbox{$\Delta T=18.1 \,^\circ$C}, which gives rise to \mbox{$h_\infty = 280$~cm} and hence a diffusive timescale \mbox{$h_\infty^2/\kappa_s = 7.7\times10^6$~s} (or 89 days). 
\item For \mbox{$T_B=-10 \,^\circ$C}, we obtain \mbox{$\Delta T=8.1 \,^\circ$C}, which gives rise to \mbox{$h_\infty = 130$~cm} and hence a diffusive timescale \mbox{$h_\infty^2/\kappa_s = 1.5\times10^6$~s} (or 18 days).
\end{enumerate}

\subsection{Equilibrium thickness} \label{sec:equil}
The ice grows and reaches an equilibrium (steady-state) thickness $\hat{h}_\infty$ at which the growth rate factor $q=0$. The temperature $\theta=\theta(\zeta)$ alone. By equation~\eqref{eq:bc_nd1_qdef}, we observe that $(k\theta')(1)=\hat{h}_\infty$. The left-hand-side of equation~\eqref{eq:temp_nd1} is zero at steady state. Thus, by integrating the right-hand-side once and applying the condition at $\zeta=1$, we obtain
\begin{equation} \label{eq:temp_steady_1}
 \left(1-\hat{S}\Delta k \frac{\theta_B-1}{\theta_B-\theta} \right)\frac{d \theta}{d \zeta} =\hat{h}_\infty,
\end{equation}
where we used equation~\eqref{eq:c_ndim} to express $k$. 
This is a first-order separable equation with two boundary conditions ($\theta(0)=0$, $\theta(1)=1$) which allows us to determine the unknown parameter $\hat{h}_\infty$.
We integrate again and apply the boundary conditions to obtain
\begin{align}\hat{h}_\infty&=1-\hat{S}\Delta k\left(\theta_B-1\right)\log \frac{\theta_B}{\theta_B-1}, \nonumber \\ &= 1-\hat{S}\Delta k\theta_0^{-1}\log (1+\theta_0), \label{eq:nd_h_inf}
\end{align}
where we used equation~\eqref{eq:theta_0} to convert from $\theta_B$ to $\theta_0$.

The equilibrium thickness calculated decreases linearly with salinity $\hat{S}$, which is driven by the thermal conductivity difference $\Delta k$. The dependence is also affected by the thermal parameter $\theta_0$. When $\theta_0\ll1$,  we note that $\theta_0^{-1}\log (1+\theta_0)\sim 1 + O(\theta_0)$. This gives $\hat{h}_\infty\approx 1-\hat{S}\Delta k$, an estimate that could be derived using a simple zero-layer type of model of ice in which the temperature field $\theta\approx \zeta$. However, when $\theta_0$ is larger, the sensitivity of equilibrium thickness to salinity is smaller. For example, when $\theta_0\approx 9.4$, the largest value considered in section~\ref{sec:parameters}, $\theta_0^{-1}\log (1+\theta_0)\approx 0.25$.

\subsection{Quasi-static approximation} \label{sec:QS}
A crucial simplification we can make to the system of equations is to neglect the explicit time-dependence $\partial \theta / \partial \tau$ in the heat equation~\eqref{eq:temp_nd1}. This is a major simplification because it reduces a partial differential equation (PDE) to an ordinary differential equation (ODE). The resulting equations are still time-dependent because the ice thickness depends on time and this affects the solution via the definition of $q$ in equation~\eqref{eq:bc_nd1_qdef}. For constant boundary conditions and when $\hat{h}\ll 1$, the quasi-static solution is an exact solution of the full PDE. This is sometimes called a similarity solution, an idea used in previous analytical studies \citep{Stefan1891,Huppert1985,Worster1986}. Thus the quasi-static approximation can be thought of as a generalization of a similarity solution.

To motivate this approximation, consider the initial phase of ice growth (from an initial thickness of zero). We scaled the equations such that $q=O(1)$ provided $\hat{S}\neq 1$, \textit{cf}.  equation~\eqref{eq:bc_nd1_qdef}. Then, by integrating equation~\eqref{eq:q_dh},
\begin{equation} \label{eq:h_early}
\hat{h} \propto \sqrt{\frac{2 \tau}{\hat{L}}}, 
\end{equation}
so the first term in the heat equation~\eqref{eq:temp_nd1} is a factor of $\tau$ smaller than the second term. Therefore, initially at least, we can guarantee that the explicit time dependence is negligible. Moreover, at later times, the system is likely to be evolving slowly anyway. We will test the practical effects of this approximation later by comparing a full solution of the PDE to the ODE (section~\ref{sec:time-non-QS}). 

Under the quasi-static approximation, the heat equation~\eqref{eq:temp_nd1} reduces to a boundary-value problem (BVP) 
 
\refstepcounter{equation}
\begin{equation*} \label{eq:temp_bvp1}
 -\frac{c \zeta q}{\hat{L}}  \theta'  =    \left(k \theta' \right)', \quad \theta(0)=0, \, \theta(1)=1,  \, q=\frac{ (k\theta')(1) -\hat{h}}{1-\hat{S}+\theta_e}. 
\eqno{(\theequation{\mathit{a},\mathit{b},\mathit{c},\mathit{d}})}
\end{equation*} 
 
Equation~(\ref{eq:temp_bvp1}\textit{a}) is a second-order ODE with an unknown parameter $q$, hence the three boundary conditions (equations \ref{eq:temp_bvp1}\textit{b,c,d}). The left-hand-side of~(\ref{eq:temp_bvp1}\textit{a}) is a pseudo-advection term associated with the change of coordinate system.

The boundary-value problem is coupled to an initial-value problem (IVP) for ice thickness (equation~\ref{eq:q_dh}). We define 
\begin{equation} \label{eq:yh_relation}
\hat{y}=\hat{h}^2 \hat{L}/2,
\end{equation}
so that the initial-value problem has the simple form
\begin{equation} \label{eq:IVP}
\frac{d\hat{y}}{d \tau} = q, \quad \hat{y}(0)=0.
\end{equation}
The coupled BVP--IVP can be solved very straightforwardly using collocation methods for the BVP \citep{Kierzenka2001} and Runge-Kutta methods for the IVP. We present an implementation based on the MATLAB bvp4c and ode45 routine, respectively (see the data availability statement for a link to the code). Similar implementations are available in the SciPy library, for example. 

\section{Initial ice growth rate} \label{sec:initial}
In this section, we analyze the early stage of sea-ice growth from an initial thickness of zero. We focus on how the growth rate depends on the salinity of sea ice and its latent heat. 

The initial ice thickness is zero ($\hat{h}=0$), so the BVP \eqref{eq:temp_bvp1} simplifies to
    
\refstepcounter{equation}
\begin{equation*} \label{eq:temp_bvp2}
 -\frac{c \zeta q_0}{\hat{L}}  \theta'  =    \left(k \theta' \right)', \quad \theta(0)=0, \, \theta(1)=1,  \, q_0=\frac{ (k\theta')(1) }{1-\hat{S}+\theta_e}, \eqno{(\theequation{\mathit{a},\mathit{b},\mathit{c},\mathit{d}})}
\end{equation*}   
  
where we introduce the notation $q_0$ for the initial value of $q$. The solution of the BVP, and hence the value of $q_0$, will depend on all the material parameters of the system (section~\ref{sec:parameters}). 
Our primary goal in this section is to analyze the role of salinity $\hat{S}$  in controlling initial ice growth. We will start by deriving analytical solutions valid when  $\hat{L}\gg 1$ (section~\ref{sec:large-L}) and $\hat{S}\ll 1$ (section~\ref{sec:initial-S}). This analytical approach elucidates the physical mechanisms at play. Then we will perform numerical calculations across the full parameter range (section~\ref{sec:initial-numerical}). 

\subsection{Analytical solution when the Stefan number is large} \label{sec:large-L}
The parameters of the system (section~\ref{sec:parameters}) suggest various approximations that simplify the analysis. We take $\Delta c = \theta_e = 0$ in this subsection and the next. In this subsection, we take $\hat{S}=0$ (fresh ice) and investigate the effect of latent heat in the limit that the Stefan number is large, $ \hat{L}\gg 1$. Given that $\hat{S}=0$, the parameters $\Delta k$ and $\theta_B$ do not affect the solution. 

The Stefan problem when $\hat{S}=0$ is very well known so we only sketch the solution. The temperature 
\begin{equation}
\theta = \frac{\textrm{erf} (C \zeta)}{\textrm{erf} (C)},
\end{equation}
where $C=\left(q_0/2\hat{L}\right)^{1/2}$ and $q_0 = (2/\sqrt{\upi}) C\exp\left(-C^2\right)/ \textrm{erf}(C)$. By eliminating $C$, we obtain an implicit algebraic expression governing $q_0(\hat{L})$. Finally, by taking the limit $\hat{L}\rightarrow \infty$ (in which $C\rightarrow 0$), we obtain the asymptotic estimate
\begin{equation} \label{eq:q0_L}
q_0  \sim 1 -\frac{1}{3} \frac{1}{\hat{L}} +\frac{7}{45} \frac{1}{\hat{L}^2} + O(\hat{L}^{-3}).
\end{equation}

\begin{figure}
  \centerline{\includegraphics{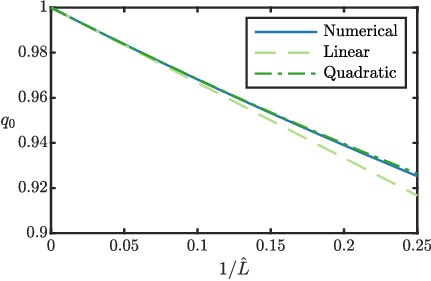}}% Images in 100% size
  \caption{Dependence of initial growth rate factor $q_0$ on the latent heat $\hat{L}$ calculated numerically and asymptotically (equation~\ref{eq:q0_L}).}
\label{fig:asymptotic_q0_L}
\end{figure}

Figure~\ref{fig:asymptotic_q0_L} shows that the asymptotic estimate is extremely close to the full numerical solution throughout the parameter range of interest. Indeed, even the leading order estimate $q_0 \sim 1 $ has an error of less than about 4\% for $\hat{L}>8$ ($1/\hat{L}<0.125$), the geophysical range of interest identified in section~\ref{sec:parameters}. 

The corresponding leading order solution $\theta = \zeta$, i.e., the temperature field is approximately linear through the ice. Thus models that assume a linear temperature profile such as the Semtner zero-layer model \citep{Semtner1976} are effective provided the latent heat is sufficiently large. 

\subsection{Analytical solution when the bulk salinity is small} \label{sec:initial-S}
We now extend the analysis to consider small but non-zero bulk salinity by investigating the limit $\hat{S}\ll 1$.  We continue to make the assumption that $ \hat{L}\gg 1$ but the solution now depends additionally on the parameters $\Delta k$ and $\theta_B$.  As we discussed in section~\ref{sec:intro}, one of the major developments in recent sea-ice modelling has been dynamically calculating the bulk salinity. But different models of bulk-salinity evolution seem to have little effect on ice thickness \citep{Griewank2013,ReesJones2014,Worster2015}. Our goal is to understand the physical reasons for this limited sensitivity and explore how this is manifested in common large-scale sea-ice models.

The key idea is to decompose the temperature field 
\begin{equation} \label{eq:decomposition}
\theta \sim \zeta + \hat{S} \tilde{\theta}(\zeta) + O(\hat{S}^2,\hat{L}^{-1}),
\end{equation}
where the perturbation temperature field $\tilde{\theta}$ satisfies a second-order BVP subject to $\tilde{\theta}(0)=\tilde{\theta}(1)=0$. This BVP is obtained by substituting equation~\eqref{eq:decomposition} into equation~\eqref{eq:temp_bvp2} and equating terms of the same order in $\hat{S}$ (i.e., we linearize the system in $\hat{S}$). 

We give full details in appendix~\ref{app:salinity}. Here we explain the main physical simplifications involved. The liquid fraction $X=\hat{S}(\theta_B-1)/(\theta_B-\theta)$, so to leading order $X$ is proportional to $\hat{S}$ and the temperature field in the denominator can be replaced by $\theta\sim\zeta$. Then the heat capacity ratio on the LHS of equation~(\ref{eq:temp_bvp2}\textit{a}) can be estimated by $c/\hat{L} \sim X/(\theta_B-\zeta)$ for similar reasons. The thermal conductivity $k=1-\Delta k X$ so there is a constant part (which must be retained), and a part that is proportional to $X$, and hence $\hat{S}$, so is retained too. Thus, it is important to consider how thermal conductivity varies with liquid fraction. Finally, the Stefan condition~(\ref{eq:temp_bvp2}\textit{d}) is linearized as follows
\begin{equation} \label{eq:q0_v1}
q_0=\frac{ (k\theta')(1) }{1-\hat{S} } \sim 1 + \hat{S}\left[1 -\Delta k +\tilde{\theta}'(1) \right]+ O(\hat{S}^2,\hat{L}^{-1}).
\end{equation}
The term in square brackets includes three terms that control the sensitivity of the growth rate of ice to its salinity. 

The first term comes from linearizing the denominator of the expression for $q_0$. Physically, this term arises from the latent heat released at the ice--ocean interface. A larger $\hat{S}$ increases the liquid fraction at the interface which reduces the latent heat liberated there which in turn increases the growth rate. That is, salty ice tends to grow quicker than fresher ice because there is less latent heat to be conducted away from the growing interface. 

The second term comes from linearizing the heat flux at the interface: considering the variation in thermal conductivity while retaining the leading order estimate $\theta'(1)\sim 1$. At the interface, $\theta=1$, so the terms involving $\theta_B$ cancel and we are left with a contribution $-\Delta k$. Physically, a higher $\hat{S}$ increases the liquid fraction which reduces the thermal conductivity because liquid water is less conductive than ice. That is, salty ice would tend to grow slower than fresher ice because it is less thermally conductive.  

The combination of the first and second terms is consistent with the prediction of a simple zero-layer type of model of ice. \citet{Worster2015} discussed the relative insensitivity of ice growth to salinity in terms of the competing effects of thermal conductivity and latent heat growth. Here, we give a clear quantification of that competition. In particular, 
\begin{equation} \label{eq:dq0_zero}
\frac{\partial q_0}{\partial \hat{S}} \approx 1 - \Delta k \approx 0.26.
\end{equation}

So the overall sensitivity of ice growth to salinity is rather weak because the latent heat and the thermal conductivity dependencies trade-off very strongly, even though individually they vary very strongly with salinity (see figure~\ref{fig:summary}). Mathematically the weak sensitivity arises because the parameter group $1 - \Delta k$ is small. Note that, from the definition of $\Delta k$, we have \mbox{$1 - \Delta k = k_l/k_s$}, so the weak sensitivity occurs in practice because $k_l$ is much smaller than $k_s$. 

Furthermore, we identify an additional term, $\tilde{\theta}'(1)$, which is the third term in the bracket of equation~\eqref{eq:q0_v1}.
This term also comes from linearizing the heat flux at the interface, this time considering the changed temperature profile $\tilde{\theta}(\zeta)$ through the ice while fixing the leading order estimate $k=1$. Calculating this term is much more involved as it requires us to solve the BVP for $\tilde{\theta}(\zeta)$. 

\begin{figure}
  \centerline{\includegraphics{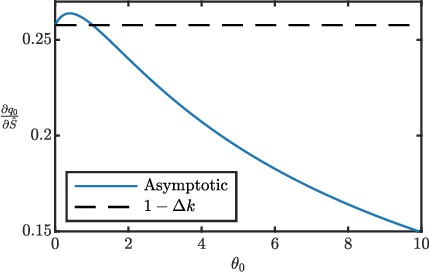}}% Images in 100% size
  \caption{Sensitivity of the initial growth rate factor to salinity at small $\hat{S}$ calculated asymptotically (equation~\ref{eq:q0_v3}). Note that the vertical axis does not begin at zero.}
\label{fig:asymptotic_dq0dS}
\end{figure}

The total effect of all these terms can be found by calculating $\tilde{\theta}'(1)$ and combining with the other terms (see appendix~\ref{app:salinity} for details). We obtain the asymptotic estimate
\begin{equation} \label{eq:q0_v2}
q_0 \sim 1+\hat{S}(\theta_B-1)\left[-2-(2\theta_B-\Delta k)\log(1-\theta_B^{-1})\right]+O(\hat{S}^2,\hat{L}^{-1}).
\end{equation}
This estimate differs from the zero-layer type estimate (equation~\ref{eq:dq0_zero}) in two main respects. It depends on $\theta_B$ whereas the previous estimate is independent of this parameter. The dependence on $\Delta k$ is also different. 
The connection between the estimates is clearer if we express equation~\eqref{eq:q0_v2} in terms of $\theta_0$ using equation~\eqref{eq:theta_0}, and then re-write as a sensitivity
\begin{align} 
\frac{\partial q_0}{\partial \hat{S}} &\sim \theta_0^{-1}\left[-2-\left(2+2\theta_0^{-1}-\Delta k\right)\log(1+\theta_0)\right]+O(\hat{S},\hat{L}^{-1}), \nonumber \\
&\sim \left[1-\Delta k \right]+O(\hat{S},\hat{L}^{-1},\theta_0), \label{eq:q0_v3}
\end{align}
where the final result follows by taking the limit $\theta_0\rightarrow 0$.
Figure~\ref{fig:asymptotic_dq0dS} shows that the sensitivity of sea-ice growth to salinity varies significantly with the cooling rate $\theta_0$ across the parameter range of interest. The prediction of equation~\eqref{eq:q0_v3} is met at small $\theta_0$. However, for strong cooling conditions when $\theta_0=10$, the growth rate factor is about 40\% smaller than expected at small $\theta_0$, for example. Given that strong cooling (large $\theta_0$) conditions are typical, this newly identified feedback mechanism associated with alteration to the thermal profile is significant.

\subsection{Numerical results} \label{sec:initial-numerical}
The asymptotic results were derived by assuming that the salinity of ice was very small $\hat{S}\ll 1$. However, in practice, we are interested in a range of salinities $0\leq \hat{S} \leq 0.8$ as discussed in section~\ref{sec:parameters}. Therefore, we solve for the initial growth rate factor $q_0$ numerically and compare with the asymptotic predictions

\begin{figure}
  \centerline{\includegraphics{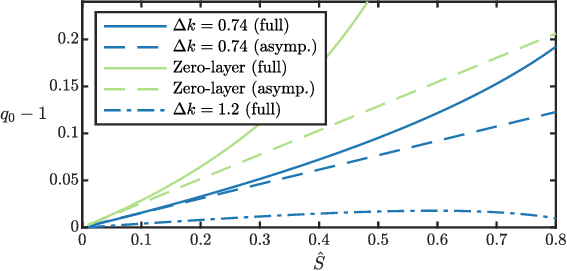}}% Images in 100% size
  \caption{Dependence of initial growth rate factor $q_0$  on the latent heat $\hat{S}$. The solid blue curve corresponds to the full numerical solution with our best estimate of the value of $\Delta k$. This approaches the asymptotic prediction (abbreviated `asymp.' in the legend) as $\hat{S}\rightarrow 0$ (blue dashed curve). The dot-dashed curve shows results for a larger value of $\Delta k$. Lighter, green curves denote the predictions of zero-layer models (solid curve shows the full model while the dashed curve is its asymptotic limit as $\hat{S}\rightarrow 0$). This figure was computed with $\hat{L}=10^{3}$, $\Delta c=\theta_e=0$ to enable direct comparison with the asymptotic theory. These simplifications are tested in figure~\ref{fig:pardep_q0}. }
\label{fig:asymptotic_q0_S}
\end{figure}

Figure~\ref{fig:asymptotic_q0_S} shows the dependence of the initial growth rate factor $q_0$ on salinity $\hat{S}$ according to various different models. The darker blue curves are the main results from our quasi-static model, with $\Delta k =0.74$, which is the best estimate for this parameter (section~\ref{sec:parameters}).  The corresponding asymptotic predictions agree with the numerical results in the limit of small $\hat{S}$ and are a reasonable approximation for $\hat{S}\lesssim 0.4$. In practice, this corresponds to a dimensional salinity of about 14 ppt, so would be relevant to all but the earliest stages of ice growth. Nevertheless, the numerical results do diverge more sharply at higher salinity, with the growth rate factor exceeding that predicted asymptotically. 

We also plot predictions from zero-layer type models in a lighter green colour, both the full calculation (solid) and the asymptotic limit from equation~\eqref{eq:dq0_zero} (dashed). These both over-predict the ice growth rate, for the reasons discussed in the previous section. Moreover, the asymptotic limit diverges markedly from the full zero-layer calculation and is a poor approximation for $\hat{S}\gtrsim 0.1$. So even though the asymptotic zero-layer model appears to be a good model, this is coincidental. 

The above calculations, including the zero-layer results, were all made with $\Delta k =0.74$. However, in section~\ref{sec:parameters}, we showed that $\Delta k \approx 1.2$ in MU-type models such as CICE/Icepack \citep{CICE2024,icepack2024}. So we also plot the numerical solution for the higher value of $\Delta k=1.2$. It has a much weaker dependence on $\hat{S}$ because the reduction in growth rate caused by thermal conductivity variation is made stronger. 

In summary, all the models we presented agree with the general idea that the sensitivity to salinity is rather weak. The zero-layer model over-predicts the sensitivity while the high-$\Delta k$ model implemented as the default parameter choice in CICE/Icepack tends to under-predict the sensitivity. 

\begin{figure}
  \centerline{\includegraphics{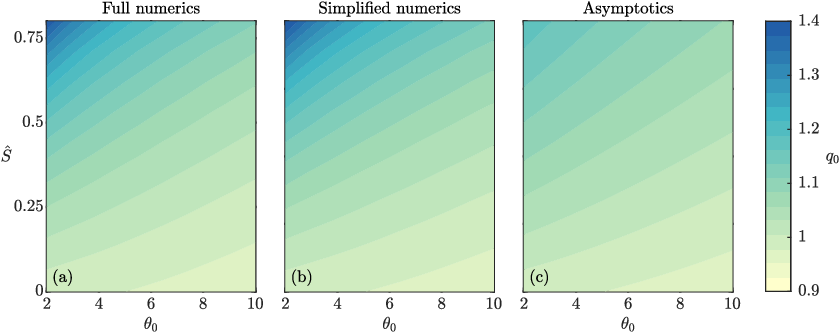}}% Images in 100% size
  \caption{Dependence of the initial growth rate factor on external parameters. (a) Full numerical results with standard material parameter values. (b) Simplified numerical results ($\Delta c = 0$ and $\theta_e=0$). (c) Asymptotic results based on the combination of equation~\eqref{eq:q0_L} for the dependence on $\hat{L}$ [where $\hat{L}=\hat{L}_0 \theta_0^{-1}$ from equation~\eqref{eq:Lhat-L0}], and equation~\eqref{eq:q0_v2} for the dependence on $\hat{S}$. Note that we only plot $\theta_0\geq 2$ to focus on the parameter regime of geophysical interest. For smaller values $\theta_0\geq 2$, $q_0$ increases rapidly in both sets of numerical calculations.   }
\label{fig:pardep_q0}
\end{figure}

In figure~\ref{fig:pardep_q0}, we show more completely the dependence of the growth rate factor on the environmentally varying parameters of the system ($\theta_0,\hat{S}$). Comparing panels (a) and (c), we see that the basic trends found numerically are consistent with the asymptotic analysis across the whole parameter space explored. Comparing panels (a) and (b) shows that relaxing the assumptions that $\Delta c = 0$ and $\theta_e=0$ makes almost no difference to results, so this was a good assumption across the full parameter regime considered. Thus the asymptotic analysis (panel c) and the physical mechanisms identified using it (section~\ref{sec:initial-S}) explain why the initial growth rate varies very weakly with salinity.

To conclude, the weak effect on relative growth rate is not only caused by the relatively low salinity $\hat{S}$ (and hence low liquid fraction). For example, even a relatively high salinity ($\hat{S}=0.5$, 50\% of the ocean seawater salinity) corresponds to $q_0-1\approx 0.1$, which is about a 10\% difference in growth rate factor. Thus the physical mechanisms, first the trade-off in thermal properties (conductivity/latent heat capacity) and second the feedback on temperature profile,  (analyzed asymptotically in section~\ref{sec:initial-S}) are crucial to understanding the role of salinity.  In the next section, we will investigate how this feeds through to the evolution of ice thickness. 

\section{Later-stage ice growth rate and thickness evolution} \label{sec:later-stage}
As the ice continues to grow, the heat flux from the ocean gradually plays a greater role slowing the ice growth until the thickness reaches a steady state (assuming constant forcing, something we will revisit in section~\ref{sec:time-dependent}). In this section, we calculate how the growth rate factor $q$ depends on the ice thickness and then use this to determine the evolution of the ice thickness. We show that the sensitivity of the growth rate to salinity reverses sign, so saltier ice grows slower. 

\subsection{Effect of ice thickness on the growth rate factor} \label{sec:ice-thickness-growth-rate}
Under the quasi-static approximation (section~\ref{sec:QS}), the problem of calculating the ice growth rate factor does not depend on the evolution of the ice thickness, only the thickness at a given time. So we now solve the full BVP~\eqref{eq:temp_bvp1} and denote the growth rate factor  $q(\hat{h})$. This notation suppresses the dependence on all the material parameters of the system. Note that $q_0=q(\hat{h}=0)$. However, we can not simply take our solutions for $q_0$ and find $q$ by using the Stefan condition (\ref{eq:temp_bvp1}\textit{d}), because $q$ also appears in the heat equation (\ref{eq:temp_bvp1}\textit{a}).

\begin{figure}
  \centerline{\includegraphics{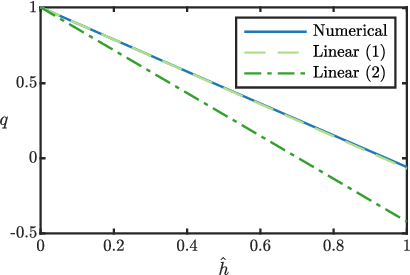}}% Images in 100% size
  \caption{  Dependence of growth rate factor $q$ on the thickness $\hat{h}$ for $\hat{S}=0.3$. The solid blue curve corresponds to the full numerical solution. The dashed light green line shows a first linear approximation given by equation~\eqref{eq:qh_approx}. This agrees very well with the full numerical solution. The dot-dashed darker green line shows a second alternative linear approximation $q=q_0 -\hat{h}/(1-\hat{S}+\theta_e)$ which does not agree with the numerical calculation. }
\label{fig:pardep_qh}
\end{figure}

Numerically, we observe that the $q(\hat{h})$ varies approximately linearly with $\hat{h}$. Figure~\ref{fig:pardep_qh} shows that this approximation holds very well across the full range of $\hat{h} \in [0,1]$ for parameter values that span the range of interest. While a linear dependence on $\hat{h}$ might be anticipated from equation (\ref{eq:temp_bvp1}\textit{d}), the slope is not consistent with $(1-\hat{S}+\theta_e)^{-1}$. This is because the temperature profile itself also changes with $\hat{h}$. 

Motivated by the observed linear trend in figure~\ref{fig:pardep_qh}, we introduce this linear approximation
\begin{equation} \label{eq:qh_approx}
q(\hat{h}) \approx q_0 - q_1 \hat{h},
\end{equation}
where $q_0$ and $q_1$ depend on the system parameters but not $\hat{h}$. We calculate $q_1$ by computing $q$ at a very small value of $\hat{h}$ to input, along with $q_0$, into a finite difference approximation of the (negative) slope at $\hat{h}=0$.

\begin{figure}
  \centerline{\includegraphics{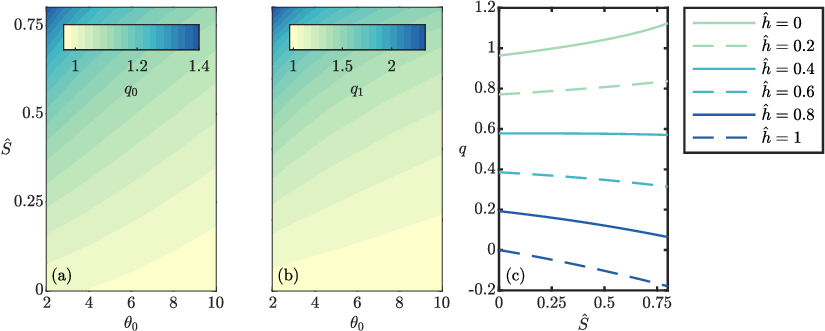}}% Images in 100% size
  \caption{Dependence of the growth rate factor on external parameters. Panels (a) and (b) show how the constant and linear terms in  equation~\eqref{eq:qh_approx}, respectively, depend on external parameters. These parameters are the same as the `full numerics' parameters of figure~\ref{fig:pardep_q0}. Panel (c) shows the resulting sensitivity of the growth rate factor to salinity at increasing ice thickness. A negative $q$ corresponds to thickness decaying towards its equilibrium state.}
\label{fig:pardep_q1}
\end{figure}

Figure~\ref{fig:pardep_q1} shows that $q_1$ (panel b) has a similar parametric dependence as $q_0$ (panel a). In particular, $q_1$ increases strongly with salinity.
Indeed the dependence of $q_1$ on salinity is stronger than that of $q_0$. The practical implication is that $q(\hat{h})$ can decrease with salinity rather than increase. Figure~\ref{fig:pardep_q1}(c) shows that at small $\hat{h}\lesssim 0.4$, the growth rate factor increases with $\hat{S}$ (consistent with the initial growth rate sensitivity found in section~\ref{sec:initial-S}) but for larger $\hat{h}\gtrsim 0.4$, the trend reverses. This reversed trend is consistent with the fact that the equilibrium (steady state) thickness decreases with salinity (see equation~\eqref{eq:nd_h_inf} in section~\ref{sec:equil}). 

Physically the reversal in sensitivity to salinity can be understood by considering the changing relative importance of the different processes that control the sensitivity to salinity. As the ice grows, the growth rate reduces, which reduces the importance of the latent heat sensitivity to salinity (recall that saltier ice grows faster as less latent heat is released by saltier ice). The turbulent ocean heat flux (a constant, independent of salinity) becomes relatively more important. The conductive heat flux reduces, but at a slower rate than the latent heat production reduces. Indeed, as the thickness approaches a steady state (section~\ref{sec:equil}), the latent heat production is negligible and there is a balance between turbulent ocean heat flux and conductive heat flux through the ice. Thus the fact that the saltier ice is less thermally conductive eventually dominates the overall sensitivity to salinity, such that saltier ice grows slower beyond $\hat{h}\gtrsim 0.4$.

\subsection{Ice thickness evolution} \label{sec:ice-thickness}
As the ice grows $\hat{h}$ increases, so while \textit{initially} saltier ice grows faster, \textit{subsequently} saltier ice grows slower. This is a further reason that the sensitivity of ice growth to ice salinity is small.  
We next investigate the net result by calculating the thickness evolution using equation~\eqref{eq:IVP}. If the growth rate factor has the simple linear form given by equation~\eqref{eq:qh_approx}, then the evolution equation becomes
\begin{equation}
\frac{d\hat{y}}{d\tau} = q_0 - \left(2 q_1^2/\hat{L}\right)^{1/2} \hat{y}^{1/2}, \qquad \hat{y}(0)=0,
\end{equation}
where we used equation~\eqref{eq:yh_relation} to convert between scaled thickness $\hat{h}$ and scaled squared thickness $\hat{y}$.
This is a separable equation and the analytical solution can be written in terms of the product-logarithm function (sometimes called the Lambert $W$-function). This function is defined as the root $w=W$ of $we^w=z$ (the principle value, which satisfies $W\geq -1$, is the relevant root here). 

We express the analytical solution in terms of the thickness 
\begin{equation} \label{eq:ht_approx}
\hat{h}=
\frac{q_0}{q_1} \left\{1 + W\left[-\exp \left(\frac{-\tau  q_1^2}{\hat{L} q_0} -1\right) \right] \right\}.
\end{equation}
To my knowledge, this expression is a new analytical estimate of the growth trajectory of sea ice. 
A related expression is given in an implicit form in \citet{Worster2000} and in the PhD thesis \citet{Notz2005} but these earlier expressions are based on a zero-layer model for ice growth. 

\begin{figure}
  \centerline{\includegraphics{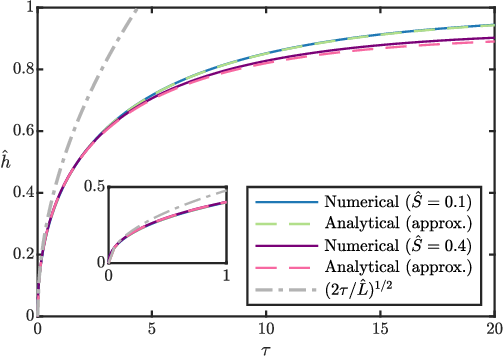}}% Images in 100% size
  \caption{Ice thickness evolution calculated according to the numerical model at two different salinities. Dashed curves show the corresponding analytical approximation from equation~\eqref{eq:ht_approx}, which are extremely close to the numerical curves. The dot-dashed curve shows an approximation to the initial growth based on $q_0\approx 1$ from equation~\ref{eq:q0_L}. The inset shows the same data over the initial phase of growth. }
\label{fig:ht_steady}
\end{figure}

Figure~\ref{fig:ht_steady} shows that the thickness is initially proportional to $\tau^{1/2}$ (see inset), as is typical for Stefan problems \citep{Worster2000}. This comes from taking $q_0\approx 1$ from equation~\ref{eq:q0_v2}, which is valid at small $\hat{S}$ and large $\hat{L}$. 
However, the thickness diverges markedly from this initial growth even by $\tau=1$. 
The growth slows and approaches a steady state as $\tau\rightarrow \infty$.
The new analytical solution from equation~\eqref{eq:ht_approx} is an excellent approximation to the full numerical solution across the full range of times considered. 
We plot results for two values of the ice salinity as examples. The initial growth of the saltier ice is slightly faster (but the curves are virtually indistinguishable). At late times, the saltier ice reaches a smaller thickness, consistent with the growth rate factor being a decreasing function of salinity once $\hat{h}\gtrsim 0.4$, as shown in figure~\ref{fig:pardep_q1}(\textit{c}). Thus the competing sensitivity of growth to salinity at small versus large thickness is an additional reason why the overall ice thickness is relatively insensitive to salinity. For practical purposes, the calculations with different salinities are virtually indistinguishable up to about $\tau\approx 5$, which dimensionally corresponds to a period of at least about 100 days (based on the \mbox{$T_B=-10 \,^\circ$C} conversion in section~\ref{sec:par-dim}). This is a remarkable degree of insensitivity in the context of the significant variation in the thermal properties of ice between these salinities.  

\section{Effect of time-dependent forcing and the validity of the quasi-static approximation} \label{sec:time-dependent}
Sea ice grows in highly variable environmental conditions. In this section, we progressively relax our assumptions of conditions being fixed. 
First, we consider the effect of variable atmospheric conditions (section~\ref{sec:time-atm}).
Second, we consider the effect of variable salinity  (section~\ref{sec:time-salinity}). 
Third, and most significantly, we relax the quasi-static approximation and compare solutions based on solving the full PDE system to solutions based on our quasi-static approximation (section~\ref{sec:time-non-QS}).

\subsection{Time-dependent atmospheric temperature} \label{sec:time-atm}
Variable atmospheric conditions are an important factor in sea-ice growth. We adjust the boundary condition at the ice--atmosphere interface. We retain the same non-dimensionalization but interpret $T_B$ as the long-term average boundary temperature. In dimensionless variables, the only change to the model is that
\begin{equation}
\theta(0)=\hat{g}(\tau),
\end{equation}
where $\hat{g}(\tau)$ represents the time-dependent boundary temperature. Within the quasi-static framework, the growth rate factor only depends on the value of $\hat{g}$ at a particular time, so we denote this dependence $q(\hat{h}, \hat{g})$, again suppressing the dependence on material properties.

\begin{figure}
  \centerline{\includegraphics{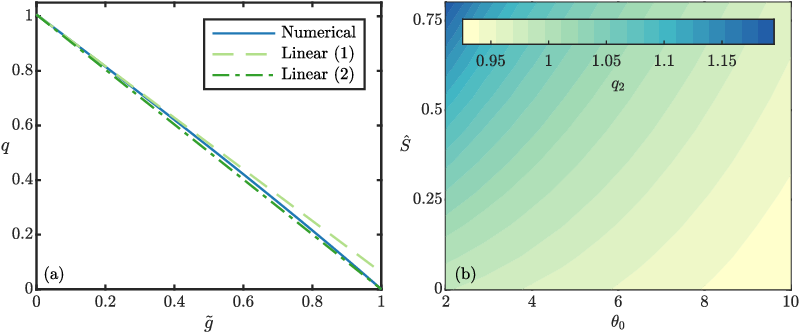}}% Images in 100% size
  \caption{Panel (a) shows the approximately linear dependence of the growth rate factor on $\hat{g}$ for $\hat{h}=0$ and $\hat{S}=0.3$. The first linear model (1) is based on numerical estimation of $q_2$ at small $\hat{g}$ as given by equation~\eqref{eq:qhg_approx}. The second linear model (2), based on a zero-layer model, is given by equation~\eqref{eq:qhg_zero}. Panel (b) shows the full parameter dependence of the slope $q_2$. }
\label{fig:pardep_q2}
\end{figure}

Figure~\ref{fig:pardep_q2}(a) shows that $q$ decreases approximately linearly with $\hat{g}$. 
A higher value of $\hat{g}$ corresponds to a higher (warmer) atmospheric temperature so slower ice growth.
We introduce the approximation 
\begin{equation} \label{eq:qhg_approx}
q(\hat{h}, \hat{g})\approx q_0 - q_1 \hat{h}-q_2 \hat{g},
\end{equation}
where $q_2$ is estimated in an analogous way to $q_1$. A naive, and rather good estimate of $q_2$ can be obtained by substituting the zero-layer estimate $\theta'(1)=1-\hat{g}$ into equation~(\ref{eq:temp_bvp1}\textit{d}), which gives
\begin{equation}\label{eq:qhg_zero}
q(0, \hat{g})\approx q_0(1- \hat{g}),
\end{equation}
or $q_2\approx q_0$.
Figure~\ref{fig:pardep_q2}(b) shows that $q_2$ varies only weakly with the external parameters and does so in a similar pattern to $q_0$, although the approximation $q_2\approx q_0$ is not valid uniformly (compare figure~\ref{fig:pardep_q0}a). 

We next consider the effect on the trajectory of ice thickness by solving the IVP~\eqref{eq:IVP} that governs its evolution. The full equation requires numerical solution. However, if we adopt the linearization~\eqref{eq:qhg_approx} and take $\hat{g} = \Delta g \hat{g}(\tau)$ where $\Delta g\ll 1$, then we find
\begin{equation}
\frac{d\hat{y}}{d\tau} = q_0 - q_1 \left(2/\hat{L}\right)^{1/2} \hat{y}^{1/2}-q_2 \Delta g \hat{g}(\tau), \qquad \hat{y}(0)=0.
\end{equation}
This equation can be linearized about the solution for $\Delta g=0$ (i.e., the solution we obtained in section~\ref{sec:ice-thickness}). We let $\hat{y}=\hat{y}_0 + \Delta g \hat{y}_1$, then 
\begin{equation}
\frac{d\hat{y}_1}{d\tau} = -q_2 \hat{g}(\tau) +q_1  \left(\frac{1}{2\hat{L}}\right)^{1/2}  \frac{\hat{y}_1}{ \hat{y}_0^{1/2}} +O(\Delta g) , \qquad \hat{y}_1(0)=0.
\end{equation}
Note that the quotient is well-behaved in the limit that $\tau\rightarrow 0$. Both the numerator and denominator tend to zero, but by l'H\^{o}pital's rule,
\begin{equation}
\lim_{\tau \rightarrow 0}\frac{\hat{y_1}}{ \hat{y}_0^{1/2}} =\lim_{\tau \rightarrow 0}\frac{\hat{y}_1'}{\tfrac{1}{2} \hat{y}_0' \hat{y}_0^{-1/2}}  =\lim_{\tau \rightarrow 0}  \frac{-q_2\hat{g}(0) }{\tfrac{1}{2} q_0 } \hat{y}_0^{1/2}  = 0,
\end{equation} 
where, in the second expression, a prime denotes a derivative with respect to $\tau$.
For large latent heat $\hat{L}\gg 1$, the evolution equation for $\hat{y}$ has a simple form:
\begin{equation} \label{eq:yhat_linear}
\frac{d\hat{y}_1}{d\tau} = -q_2 \hat{g}(\tau) +O(\Delta g,\hat{L}^{-{1/2}}) , \qquad \hat{y}_1(0)=0.
\end{equation}

We use these ideas to propose three solutions to the ice-thickness evolution equation:
\begin{enumerate}
\item A numerical solution to the IVP with the full $q(\hat{h}, \hat{g})$.  
\item A numerical solution to the IVP with the linearized form~\eqref{eq:qhg_approx}.  
\item An analytical solution to the IVP based on equation~\eqref{eq:yhat_linear}.
\end{enumerate}

\begin{figure}
  \centerline{\includegraphics{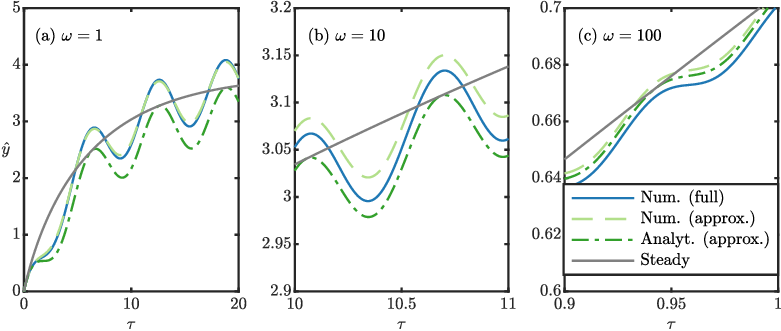}}% Images in 100% size
  \caption{The evolution of ice plotted in terms of the squared thickness scale $\hat{y}=\hat{h}^2 \hat{L}/2$ under sinusoidal atmospheric temperature variation $\hat{g}=\Delta g \sin(\omega \tau)$ for $\hat{S}=0.3$. In all panels $\Delta g=0.5$ to allow us to test the success of the approximate solutions beyond the  $\Delta g\ll 1$ limit in which they are derived. The panels show different forcing frequencies: (a) low frequency, (b) intermediate frequency and (c) high frequency. For the latter plots, we restrict the $\tau$ axis range to show a representative part of the solution. Solid blue curve shows the full numerical solution, referred to as (i) in the main text. Dashed green curves show an approximate numerical solution (ii). Dot-dashed dark green curves show an analytical approximation (iii). Solid grey curves show the steady solution equivalent to equation~\eqref{eq:ht_approx}.}
\label{fig:yt_sinusoidal}
\end{figure}

Figure~\ref{fig:yt_sinusoidal} shows the ice-thickness evolution for a sinusoidal forcing $\hat{g}=\Delta g \sin(\omega \tau)$, where $\omega$ is the angular frequency. While realistic forcing would contain a large spectrum of frequencies, using a simple sinusoidal forcing allows us to test the success of the various approximations we introduced as a function of frequency. For intermediate and high frequency ($\omega =\{10,100\}$, corresponding, roughly, to sub-fortnightly time periods) all the approximations we introduced agree very well with the full numerical solution. (To be more precise, for the \mbox{$T_B=-10 \,^\circ$C} conversion in section~\ref{sec:par-dim}, $\omega=10$ corresponds to a period of 13 days.) For such frequencies, the basic behaviour can be most easily seen by integrating equation~\eqref{eq:yhat_linear} which gives
\begin{equation}
\hat{y}_1 \approx \frac{q_2 \Delta g  }{\omega} \left[\cos(\omega t)-1\right].
\end{equation}
The growth rate is anti-phase with the forcing (because a peak in boundary temperature corresponds to a trough in growth rate). Then $\hat{y}$ lags the peak in growth rate by a quarter of a cycle. 
For longer-term variability (e.g., $\omega=1$), the approximations (i/ii) agree well. However, there is a more significant departure from (iii). This is expected because $\hat{y}\propto \omega^{-1}$, so the $O(\hat{L}^{-1/2})$ term neglected in deriving equation~\eqref{eq:yhat_linear} will be smaller at high frequency than at low frequency. Similar behaviour has been observed in experimental systems with analogous behaviour \citep{Ding2019}.

\subsection{Time-dependent salinity} \label{sec:time-salinity}
We next investigate the effect of time-dependent salinity on ice growth. The evolution of the salinity profile is complex. Here, we investigate a simple prescribed time-dependent salinity to assess whether or not it has a large effect on ice growth. In particular, we let
\begin{equation} \label{eq:S_time_dep}
\hat{S}(\tau)=\hat{S}_2 + \left( \hat{S}_1 - \hat{S}_2 \right) \exp(-\tau/\tau_d),
\end{equation}
where $\hat{S}_1$ is the initial salinity, $\hat{S}_2$ is the late-time salinity and $\tau_d$ is the desalination timescale over which the salinity evolves. 
This type of exponential relaxation has been used previously by \citet{Vancoppenolle2009a} and then used as `slow mode' of gravity drainage by \citet{Turner2013}. While it does not arise from any particular physical model of desalination, it is a simple functional form to explore. 

\begin{figure}
  \centerline{\includegraphics{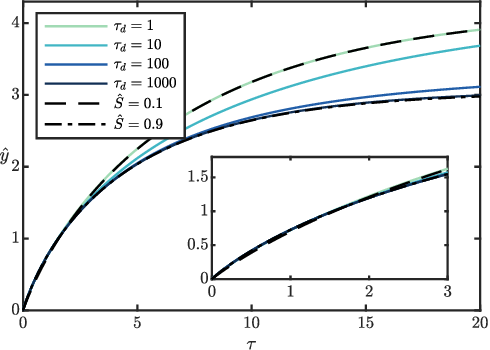}}% Images in 100% size
  \caption{Ice growth plotted in terms of the squared thickness scale $\hat{y}=\hat{h}^2 \hat{L}/2$ with time-dependent prescribed salinity according to equation~\eqref{eq:S_time_dep} with initial salinity $\hat{S}_1=0.9$ and late-time salinity $\hat{S}_2=0.1$. The solid curves correspond to different desalination timescales. The dot-dashed and dashed curves are constant-salinity calculations corresponding to the initial and late-time salinity, respectively. The inset shows the early time behaviour in which all the curves are almost indistinguishable. Close inspection shows that the $\hat{S}=0.1$ dashed curve is lowest early on, while the $\hat{S}=0.9$ dot-dashed curve is lowest at late times.   }
\label{fig:yt_varS}
\end{figure}

Figure~\ref{fig:yt_varS} compares the thickness evolution with a constant salinity (corresponding to either the initial or late-time salinity) with that with an evolving salinity. We look at an extreme example with a high initial salinity $\hat{S}_1=0.9$ and low late-time salinity $\hat{S}_2=0.1$ in order to consider a significant salinity drop. With a relatively rapid desalination timescale $\tau_d\leq 1$, the thickness is extremely close to a fixed salinity model with the prescribed late-time salinity. For longer desalination timescales, the salinity remains higher for longer. The thickness increases more slowly. For extremely long desalination timescales $\tau_d =100,1000$, much longer than the maximum $\tau$ considered, the thickness evolution is close to a fixed salinity with the prescribed initial salinity. 

In practice, we expect ice to desalinate significantly within the 18 days of growth that corresponds in dimensionless units to $\tau=1$ (for the \mbox{$T_B=-10 \,^\circ$C} conversion in section~\ref{sec:par-dim}, and even lower $\tau$ for colder $T_B$), so fixed-salinity models with the late-time salinity are likely to be excellent approximations. A 20-day desalination timescale was previously used for the winter by \citet{Vancoppenolle2009a}. Moreover, all the models give extremely similar results (see the inset plot looking up to $\tau=3$ which corresponds approximately to the first two months of growth). Later stages of growth are sensitive to the later-time salinity value. 

Although the quasi-static model we developed cannot consider the full dynamic evolution of salinity, it could in principle be extended to consider salinity profiles of the separable form
\begin{equation}
\hat{S} = r(\tau) s(\zeta),
\end{equation}
where $r$ and $s$ are given functions. 
However, caution is needed as the assumptions used to justify the heat equation~\eqref{eq:temp_1} would not be formally valid, as described in appendix~\ref{app:heat}.

\subsection{Non-quasi-static effects} \label{sec:time-non-QS}
Finally, and most importantly, we consider the limitations of the major simplification introduced in this study, the quasi-static approximation (section~\ref{sec:QS}). For fixed boundary conditions and an initial temperature profile consistent with that calculated within the quasi-static approximations,  the full PDE solution is identical to the quasi-static (QS) solution. To provide a strong test of the QS solution, we perform experiments with a step change in the atmospheric boundary temperature at a given time $\tau_s$. In dimensional units, we switch from \mbox{$T_B=-10 \,^\circ$C} to \mbox{$T_B=-20 \,^\circ$C}. A step change is a more challenging test than the sinusoidal forcing considered in section~\ref{sec:time-atm}.
The salinity value used in these experiments was $\hat{S}\approx 0.56$, corresponding to 20~ppt. The non-dimensionalization of the temperature is based on the initial $T_B$ which gives $\theta_0\approx 4.2$ and $\hat{L}\approx 20$.

\begin{figure}
  \centerline{\includegraphics{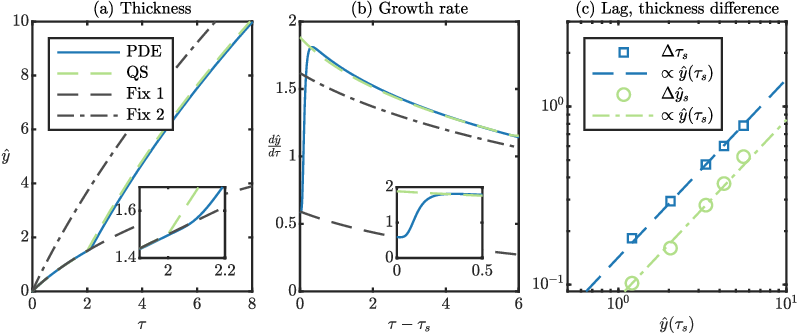}}% Images in 100% size
  \caption{Ice growth calculated by the full PDE compared against the quasi-static approximation (QS). Panel (a) shows the evolution of $\hat{y}=\hat{h}^2 \hat{L}/2$ for the PDE and QS models, as well as for two fixed temperature calculations (numbered 1: fixed at the initial atmospheric temperature; 2: fixed at the temperature after the switch occurs). The switch occurs at $\tau_s=2$. Panel (b) shows the growth rate for the same example as panel (a). Insets of both panels highlight the behaviour around the switching time. Such experiments are repeated at a series of switching times. Panel (c) shows the results of a series of experiments with $\tau_s=1,2,4,6,10$. It shows the lag $\Delta \tau_s$ before the growth rate of the PDE catches up with that of the QS model. It also shows the maximum absolute difference in $\hat{y}$ after the switching time between the PDE and QS models denoted $\Delta y_s$. Linear fits to the data are shown.}
\label{fig:yt_PDE}
\end{figure}

Figure~\ref{fig:yt_PDE}(a,b) shows an example with a switching time $\tau_s=2$. This corresponds to a dimensional time of 36~days, since we non-dimensionalize with respect to the initial value \mbox{$T_B=-10 \,^\circ$C}, which corresponds to a diffusive timescale of 18~days (section~\ref{sec:par-dim}).  The PDE and QS models agree extremely closely. For reference, fixed atmospheric temperature models with the before and after switch temperatures are also plotted. After the switch, the atmospheric temperature is much colder so the ice thickness increases more rapidly. In the QS model, the growth rate increases instantaneously. In the full PDE calculations, the growth rate increases rapidly (but not instantaneously) before catching up with the QS model over a switching lag time $\Delta \tau_s$. After this time, the growth rate of the PDE model is slightly faster than the QS because the thickness is slightly smaller. The absolute differences between the PDE and QS models in terms of thickness are very small. 

The switching lag depends on the thickness around the time the switch occurs.  Figure~\ref{fig:yt_PDE}(c) shows the results of a series of experiments at different switching times. We observe that the lag is proportional to the value of the square of the ice thickness at the switching time, i.e.,  $\Delta \tau_s \propto \hat{y}(\tau_s)$. This is because the lag corresponds to the time it takes for the cooling at the ice--atmosphere interface to diffuse across the full depth of the ice to the ice--ocean interface. A similar diffusive lag was observed experimentally by \citet{Ding2019}. In our dimensionless units, $\Delta \tau_s \propto [\hat{h}(\tau_s)]^2  \propto \hat{y}(\tau_s)$. During this period of slower growth for the PDE model, slightly less ice grows. We denote the maximum absolute difference in $\hat{y}$ between the PDE and QS models as $\Delta y_s$. We observe that $\Delta \hat{y}_s \propto  \Delta \tau_s \propto \hat{y}(\tau_s)$. The first proportionality assumes that $d\hat{y}/d\tau$ is approximately independent of switching time. While this is only an approximation, across the range considered in figure~\ref{fig:yt_PDE}(c), the discrepancy is within about $\pm 15$\%, so the dot-dashed linear fit matches the calculated $\Delta y_s$ well. Thus although the lag increases with switching time, and the square thickness change $\Delta \hat{y}_s$ increases too, the proportionate change $\Delta \hat{y}_s/\hat{y}(\tau_s)$ does not increase. Therefore, the QS model remains an effective approximation to the PDE model for any switching time. 

This was a strong test of the QS model. In practice, changes in atmospheric temperature will occur in both directions (not just a sudden cooling), which would partly offset each other. Furthermore, more realistic changes would occur more gradually than the instantaneous switch tested here. So it is reasonable to expect the QS model to perform even better under more realistic forcing scenarios. 

\section{Discussion and implications} \label{sec:implications}

This study was designed to examine and explain the observed weak sensitivity of thermodynamic sea-ice growth to salinity observed in studies with a range of fixed-salinity profiles \citep[e.g.,][]{Vancoppenolle2005} and in studies with dynamic-salinity profiles \citep[e.g.,][]{Griewank2013,ReesJones2014}. This weak sensitivity occurs despite the strong sensitivity of the thermal properties (figure~\ref{fig:summary}). There is an obvious basic reason, the sea-ice salinity is low so the solid fraction is high and the thermal properties are close to that of pure ice. However, the insensitivity persists even when the salinity is not very low (relative to the salinity of seawater). To explain the insensitivity, we identified the three main mechanisms described below. 

\subsection{Three mechanisms that explain why ice growth rate is insensitive to salinity}

\begin{description}
\item[Reason 1: trade-off between thermal conductivity and latent heat capacity. ] Saltier ice has a higher liquid fraction and hence a lower thermal conductivity, which leads it to grow more slowly. However, a higher liquid fraction also means less latent heat needs to be conducted away as the ice grows, which leads it to grow more quickly. This competition was described in \citet{Worster2015}. In this study, we give a detailed quantification and derive excellent analytical estimates for the effect (\textit{cf}. figure~\ref{fig:asymptotic_q0_S}).
\item[Reason 2: feedback on thermal profile within the ice. ] Our previous estimates of the sensitivity to salinity assumed a linear thermal profile \citep{Worster2015}, as per zero-layer Semtner models \citep{Semtner1976}. In this study, we show that feedback on the thermal profile, which produces a non-linear profile, reduces the sensitivity to salinity further and introduces a sensitivity to the cooling factor (the ratio of the temperature difference across the ice to the freezing point of seawater), as shown by figures \ref{fig:asymptotic_dq0dS} and \ref{fig:pardep_q0}.
\item[Reason 3: opposite sensitivity of thicker ice versus thinner ice. ] Saltier ice grows faster when the ice is thin, but saltier ice grows slower when the ice is thicker. This reversed sensitivity can be understood by considering the relative importance of the three mechanisms that determine the growth rate and their separate sensitivities to salinity. As well as the mechanisms described in reason 1, the turbulent ocean heat flux is a constant, independent of salinity. As the ice grows thicker, the latent heat mechanism becomes relatively less important (indeed at the steady state, equilibrium thickness, it can be neglected entirely). Thus, when the ice is sufficiently thick, the sensitivity to salinity reverses, and saltier ice grows slower (figures \ref{fig:pardep_qh} and \ref{fig:pardep_q1}). As the thickness evolves from zero towards the equilibrium thickness, the sensitivities oppose each other, leading to an even smaller net sensitivity (figure~\ref{fig:ht_steady}).
\end{description}

\subsection{Implications for large-scale sea-ice modelling} \label{sec:implications-large-scale}
Recently, there has been a move towards using `dynamic salinity' sea-ice models, as described in the introduction (section~\ref{sec:intro}). In terms of the ice-thickness evolution alone, these models behave rather similarly to fixed-salinity models throughout the winter growth season, as we and others had observed previously \citep[e.g.,][]{Griewank2013,ReesJones2014}. This study shows that this is a systematic, generic result, not merely an artefact of the particular examples calculated previously. Figure~\ref{fig:yt_varS} shows that the conclusion holds even for extreme variation in salinity across a wide range of desalination timescales. 
 
However such dynamic-salinity models are still potentially worthwhile for a range of other purposes. There is a direct connection between sea-ice desalination and the buoyancy forcing that drives ocean mixing in the polar oceans. Dynamic-salinity models will have a different time dependence of buoyancy forcing relative to fixed-salinity models \citep{Worster2015}. Furthermore, the brine motion that causes desalination also transports a wide range of biogeochemical tracers and so model differences will also affect the evolution of these species \citep{Vancoppenolle2013,Wells2019}. Differences between fixed and dynamic salinity models have been observed both in standalone sea-ice models \citep{Turner2015} and coupled Earth-system models \citep{Bailey2020}. Changing the buoyancy forcing can have large-scale impacts by changing the amount of dense bottom water being formed \citep{DuVivier2021}. Dynamic-salinity models have shown some promise in interpreting data from ice mass balance buoys \citep{Plante2024}.
 
One issue that complicates the comparison of fixed and dynamic models of salinity is the choice of parameters (material properties). While mushy-layer models are formally equivalent to Maykut--Untersteiner/Bitz--Lipscomb models, as shown by \citet{Feltham2006}, the equivalence relies on an equivalent choice of parameters. In this study, we showed that the default values in CICE/Icepack \citep{CICE2024,icepack2024} of $\beta=0.13$~W/m/ppt (which controls the sensitivity of thermal conductivity to salinity) is too large to be consistent with mushy-layer theory and would need to be reduced to $\beta = 0.082$~W/m/ppt for consistency. While it seems to be well known that the default value of $\beta$ might be problematic (it is mentioned in the documentation), this inconsistency with mushy-layer theory does not seem well known. Indeed, the recent large-scale studies cited in the previous paragraph have tended to use the `bubbly' model of thermal conductivity from \citet{Pringle2007}. \citet{Schroder2019} showed that this form for the conductivity gave the best agreement with CryoSat-2 observations of sea-ice thickness. When assessing the differences between models, it is important to distinguish structural differences (fixed versus dynamic salinity) from mere parameter choice differences. 

Finally, this study proposes an intermediate complexity class of sea-ice model, the quasi-static model. We use a changing coordinate system, transforming the system to a reference frame in which the ice is fixed. We solve a boundary-value problem (BVP) for the temperature field within the ice with no explicit time dependence but including a psuedo-advection term associated with the changed coordinate system. This new term is proportional to the ice growth rate factor. The BVP is then coupled to an initial-value problem (IVP) for the thickness evolution of time. This BVP--IVP approach is more complex than the zero-layer Semtner-type model \citep{Semtner1976} but considerably simpler than full PDE-based models such as Maykut--Untersteiner/Bitz--Lipscomb and their dynamic-salinity successors. The quasi-static model is an exact solution when the external forcing is constant, and is used here as a tool to facilitate analytical progress. We also apply it to sinusoidal atmospheric forcing with a range of periods and show that considerable analytical progress can be made in understanding the results (figure~\ref{fig:yt_sinusoidal}). Finally, we show that it represents a very good approximation to full PDE-based models even under the most difficult time-dependent atmospheric forcing scenario, a step change (figure~\ref{fig:yt_PDE}).
This type of intermediate complexity model is likely to be helpful for assessing and understanding the potential impact of any proposed changes to a large-scale model (including, for example, the parameter choices discussed above) as it can be run much more rapidly than a full-PDE model. It may also be valuable in more speculative scenarios such as in modelling potential mushy layers on icy moons \citep[e.g.,][]{Buffo2020,Buffo2021,Vance2021} where an idealized modelling approach may allow the parametric uncertainty to be explored thoroughly.

\backsection[Acknowledgements]{Insightful comments on an earlier draft by A. J. Wells and M. G. Worster helped improve this manuscript. I also appreciate the comments of two anonmyous referees.}

\backsection[Funding]{This research received no specific grant from any funding agency, commercial or not-for-profit sectors. In order to meet institutional and research funder open access requirements, any accepted manuscript arising shall be open access under a Creative Commons Attribution (CC BY) reuse licence with zero embargo.
}

\backsection[Declaration of interests]{The author reports no conflict of interest.}

\backsection[Data availability statement]{The code used to generate the results of this study is openly available in on the Zenodo repository at \url{https://zenodo.org/doi/10.5281/zenodo.13710756} \citep{ReesJones2024-code}.}

\backsection[Author ORCID]{D. W. Rees Jones, \url{https://orcid.org/0000-0001-8698-401X}}

\appendix

\section{Estimation of the relative importance of additional terms in heat equation relating to brine transport and desalination}\label{app:heat}
The heat equation derived in mushy-layer models of sea ice has additional terms relating to the evolution of bulk salinity that are not present in the heat equation~\eqref{eq:temp_1} used in this study. In particular, a more general version of this equation is 
\begin{equation} \label{eq:temp_1plus}
c \frac{\partial T}{\partial t} + {\tilde{c}} w \frac{\partial T}{\partial z}  =  \kappa_s \frac{\partial}{\partial z} \left(k \frac{\partial T}{\partial z} \right),
\end{equation}
where $w$ is the vertical brine velocity and
\begin{equation} 
{\tilde{c}} = \frac{c_l}{c_s} + \frac{L}{c_s(-T)}. 
\end{equation}
The first term in $\tilde{c}$  arises from brine advection, i.e., the heat transported by the motion of brine through the porous matrix of sea ice. 
The second term in $\tilde{c}$ arises from the desalination of the ice, which causes a reduction in liquid fraction and hence a source of latent heat. 
A full derivation is given in \citet{Feltham2006} and \citet{ReesJones2014}, for example. 

The ratio of the second to the first term in ${\tilde{c}}$ is $L/c_l(-T)$. But  \mbox{$L/c_l \approx 77 \, ^\circ$C} (section~\ref{sec:parameters}) and {$-T \gtrsim 2\, ^\circ$C}, so this ratio is much greater than 1. Physically, the heat directly transported by brine advection is always negligible compared to the contribution from latent heating. 

Equation~\eqref{eq:temp_1plus} does not determine the brine velocity $w$. In general, it will be determined by convective desalination and flushing. Determining $w$ has been one of the areas of recent development in sea-ice modelling, as discussed in section~\ref{sec:intro}. 

To estimate the order of magnitude significance of the extra term in equation~\eqref{eq:temp_1plus}, we instead use the conservation equation for bulk salinity
\begin{equation}
\frac{\partial S}{\partial t} +  w \frac{\partial C}{\partial z}  = 0,
\end{equation}
where we have neglected diffusion of salt \citep[e.g.,][]{Feltham2006,ReesJones2014}. Indeed, \citet{Wells2019} showed that salt diffusion 
had a relatively modest effect on the growth of sea ice.  
But $C$ is linearly related to $T$ through equation~\eqref{eq:C(T)}, so 
\begin{equation}
w\frac{\partial T}{\partial z} = - m  w \frac{\partial C}{\partial z}  = m \frac{\partial S}{\partial t}.
\end{equation}
This equation~\eqref{eq:temp_1plus} can be rewritten
\begin{equation} \label{eq:temp_1plus_v2}
c \frac{\partial T}{\partial t} +  {\tilde{c}}  m \frac{\partial S}{\partial t}  =  \kappa_s \frac{\partial}{\partial z} \left(k \frac{\partial T}{\partial z} \right).
\end{equation}
So if the ice salinity is constant (as we assumed throughout our calculations except in section~\ref{sec:time-salinity}), then the term involving ${\tilde{c}} $ is zero. 

For variable salinity (section~\ref{sec:time-salinity}), if we assume that timescale for salinity change $\Delta S$ occurs at a comparable rate to that of temperature $\Delta T$, then the ratio of the second to the first term on the left-hand side of equation~\eqref{eq:temp_1plus_v2} can be estimated
\begin{equation} \label{eq:temp_1plus_v2_rations}
\frac{{\tilde{c}}  m \, {\partial S}/{\partial t}  }{ c  \, {\partial T}/{\partial t} } \sim \frac{{\tilde{c}}  m \, \Delta S  }{ c  \,\Delta T }  \sim \frac{\Delta S}{S} \frac{(-T)}{\Delta T}.
\end{equation}
In deriving this estimate, we assumed that both $\tilde{c}$ and $c$ are dominated by the terms involving latent heat for the reasons discussed above. 
Both fractions at the end of equation~\eqref{eq:temp_1plus_v2_rations} are, in general, $O(1)$, so we would expect both terms on the left-hand side of equation~\eqref{eq:temp_1plus_v2} to matter.
However, within the quasi-static framework (section~\ref{sec:QS}), we map evolution in time into $(\tau,\zeta)$ coordinates and neglect any explicit dependence on $\tau$. Therefore, if salinity is taken as constant or a function of $\tau$ only, it is self-consistent to neglect the term involving $\tilde{c}$, as we have done in the main body of this study. The most complex $\tau$-dependent salinity variation considered is found in section~\ref{sec:time-salinity}, elsewhere the salinity is constant. For a more general salinity evolution, both terms are in principle the same order of magnitude. To reiterate, the contribution from desalination comes from latent heat release, not brine advection directly.

\section{Asymptotic solution of BVP for initial growth rate factor $q_0$}\label{app:salinity}
Here, we provide a more detailed derivation of the asymptotic estimates given in section~\ref{sec:initial-S}.
Under the assumptions stated in the main text ($\theta_e=0$, $\Delta c=0$, $\hat{L}^{-1}\rightarrow 0$), the full BVP (equation~\ref{eq:temp_bvp2}) can be written
\refstepcounter{equation}
  \begin{equation*} \label{eq:BVP_app}
 -\frac{c}{\hat{L}}  \zeta q_0 \theta'  =    \left(k \theta' \right)', \quad \theta(0)=0, \, \theta(1)=1,  \, q_0=\frac{ (k\theta')(1) }{1-\hat{S}}, 
\eqno{(\theequation{\mathit{a},\mathit{b},\mathit{c},\mathit{d}})}
\end{equation*}   
where the material properties are taken from equation~\eqref{eq:c_ndim},
\refstepcounter{equation}
   \begin{equation*} \label{eq:props_app}
X=     \hat{S}\frac{\theta_B-1}{\theta_B-\theta}  , \quad
     k=     1-X \Delta k , \quad
   \frac{c}{\hat{L}} =\frac{1}{\theta_B-\theta} X.     
\eqno{(\theequation{\mathit{a},\mathit{b},\mathit{c}})}
\end{equation*}   
Note that the limit $\hat{L}^{-1}\rightarrow 0$ justifies the loss of two terms in the simplified expression for $c/{\hat{L}}$ given here, although $c/{\hat{L}}$ itself must be retained. 

We then make the ansatz 
\begin{equation}
\theta\sim\zeta + \hat{S} \tilde{\theta}(\zeta) + O(\hat{S}^2),
\end{equation} and substitute this expression into the BVP. Note that ${c}/{\hat{L}}\propto X = O(\hat{S})$, so we can need only consider the leading order temperature profile ($\theta\sim\zeta$) and leading order estimate of the growth rate factor ($q_0\sim 1$) on the left-hand side of equation~(\ref{eq:BVP_app}\textit{a}). We obtain 
\begin{equation}
-\hat{S} \frac{\theta_B-1}{(\theta_B-\zeta)^2} \zeta = \left[ \left( 1 - \Delta k \hat{S}\frac{\theta_B-1}{\theta_B-\zeta}\right)\left(1+\hat{S} \tilde{\theta}' \right)\right]' + O(\hat{S}^2).
\end{equation}
By equating terms at $O(\hat{S})$, and after some algebraic manipulation, we obtain 
\begin{equation}
\tilde{\theta}'' =  \frac{\theta_B-1}{(\theta_B-\zeta)^2}  \left(  \Delta k-\zeta \right).
\end{equation}
We integrate once to obtain
\begin{equation}  \label{eq:dtheta_app}
\frac{\tilde{\theta}'}{\theta_B-1} =  \frac{\Delta k - \theta_B}{\theta_B-1}  - \log(\theta_B-\zeta) + C,
\end{equation}
where $C$ is constant. 
We integrate again to obtain
\begin{equation}
\frac{\tilde{\theta}}{\theta_B-1} =    \left(  2\theta_B - \Delta k-\zeta \right) \log(\theta_B-\zeta) + (1+C)\zeta + D ,
\end{equation}
where $D$ is constant. 
$D$ is determined by the boundary condition (\ref{eq:BVP_app}\textit{b}):
\begin{equation}
D=-(2 \theta_B-\Delta k) \log(\theta_B).
\end{equation}
Then $C$ is determined by the boundary condition (\ref{eq:BVP_app}\textit{c}):
\begin{equation} \label{eq:C_app}
C=- (2 \theta_B-\Delta k-1) \log(\theta_B-1) -1 + (2 \theta_B-\Delta k) \log(\theta_B).
\end{equation}
To calculate the (extra contribution to the) temperature gradient at the ice--ocean interface, we substitute equation~\eqref{eq:C_app} into equation~\eqref{eq:dtheta_app} at evaluate at $\zeta=1$ and obtain, after simplification,
\begin{equation} \label{eq:dtheta1_app}
{\tilde{\theta}}'(1) = 1+\Delta k -2\theta_B -(\theta_B-1)(2\theta_B-\Delta k)\log(1-\theta_B^{-1}).
\end{equation}
Finally, by expanding the final boundary condition~(\ref{eq:BVP_app}\textit{d}),
\begin{equation} \label{eq:q0_v1_app}
q_0  \sim 1 + \hat{S}\left[1 -\Delta k +\tilde{\theta}'(1) \right]+ O(\hat{S}^2),
\end{equation}
and combining with equation~\eqref{eq:dtheta1_app}, we obtain the result 
\begin{equation} \label{eq:q0_v2_app}
q_0 \sim 1+\hat{S}(\theta_B-1)\left[-2-(2\theta_B-\Delta k)\log(1-\theta_B^{-1})\right]+O(\hat{S}^2).
\end{equation}
This result is reported as equation~\eqref{eq:q0_v2} in the main text where we present numerical evidence (figure~\ref{fig:asymptotic_q0_S}) that this asymptotic limit holds well even up to moderate values of $\hat{S}$.

\bibliographystyle{jfm}
\bibliography{jfm}

\end{document}